%% file: main.tex
\documentclass[sigconf]{acmart}
\usepackage{siunitx}
\usepackage{multirow}
\usepackage{booktabs}  
\usepackage{array} 
\usepackage{colortbl}
\usepackage{xcolor}
\usepackage{tikz}
\usepackage{pgfplots}
\pgfplotsset{compat=1.18}
\makeatletter
\renewcommand\thesubsubsection{\thesubsection.\@arabic\c@subsubsection}
\makeatother
\AtBeginDocument{%
  }

\setcopyright{acmlicensed}
\copyrightyear{2025}
\acmYear{2025}
\acmDOI{XXXXXXX.XXXXXXX}
\acmConference[Conference acronym 'XX]{Make sure to enter the correct
  conference title from your rights confirmation email}{June 03--05,
  2018}{Woodstock, NY}
\acmISBN{978-1-4503-XXXX-X/2018/06}

\begin{document}
\copyrightyear{2026}
\acmYear{2026}
\setcopyright{othergov}
\acmConference[SCA/HPCAsia 2026]{SCA/HPCAsia 2026: Supercomputing Asia and International Conference on High Performance Computing in Asia Pacific Region}{January 26--29, 2026}{Osaka, Japan}
\acmBooktitle{SCA/HPCAsia 2026: Supercomputing Asia and International Conference on High Performance Computing in Asia Pacific Region (SCA/HPCAsia 2026), January 26--29, 2026, Osaka, Japan}
\acmPrice{}
\acmDOI{10.1145/3773656.3773665}
\acmISBN{979-8-4007-2067-3/2026/01}
\title{ROIX-Comp: Optimizing X-ray Computed Tomography Imaging Strategy for Data Reduction and Reconstruction}

\author{Amarjit Singh}
\email{amarjitsingh@riken.jp}
\orcid{0000-0002-7081-7329}
\affiliation{%
  \institution{RIKEN (R-CCS)}
  \city{Kobe}
  \state{Hyogo}
  \country{Japan}
}

\author{Kento Sato}
\email{kento.sato@riken.jp}
\orcid{}
\affiliation{%
  \institution{RIKEN (R-CCS)}
  \city{Kobe}
  \state{Hyogo}
  \country{Japan}
}

\author{Kohei Yoshida}
\email{yoshida@hpc.is.uec.ac.jp}
\orcid{0000-0001-7639-899X}
\affiliation{%
  \institution{The University of Electro-Communications, RIKEN (R-CCS)}
  \city{Tokyo}
  \country{Japan}
}

\author{Kentaro Uesugi}
\email{ueken@spring8.or.jp}
\orcid{}
\affiliation{%
  \institution{Japan Synchrotron Radiation Research Institute}
  \city{Sayo}
  \state{Hyogo}
  \country{Japan}
}

\author{Yasumasa Joti}
\email{joti@spring8.or.jp}
\orcid{}
\affiliation{%
  \institution{RIKEN SPring-8 Center}
  \state{Hyogo}
  \country{Japan}
}

\author{Takaki Hatsui}
\email{hatsui@spring8.or.jp}
\orcid{}
\affiliation{%
  \institution{RIKEN SPring-8 Center}
  \state{Hyogo}
  \country{Japan}
}

\author{Andrès Rubio Proaño}
\email{arubio5@indoamerica.edu.ec}
\orcid{0000-0003-1270-9140}
\affiliation{%
  \institution{Centro de Investigación en Mecatrónica y Sistemas Interactivos(MIST), Universidad Tecnológica Indoamérica}
  \city{Quito}
  \state{Pichincha}
  \country{Ecuador}
}

\begin{abstract}
  
  In high-performance computing (HPC) environments, particularly in synchrotron radiation facilities, vast amounts of X-ray images are generated. Processing large-scale X-ray Computed Tomography (X-CT) datasets presents significant computational and storage challenges due to their high dimensionality and data volume. Traditional approaches often require extensive storage capacity and high transmission bandwidth, limiting real-time processing capabilities and workflow efficiency. To address these constraints, we introduce a region-of-interest (ROI)-driven extraction framework (ROIX-Comp) that intelligently compresses X-CT data by identifying and retaining only essential features. Our work reduces data volume while preserving critical information for downstream processing tasks.
  At pre-processing stage, we utilize error-bounded quantization to reduce the amount of data to be processed and therefore improve computational efficencies. 
  At the compression stage, our methodology combines object extraction with multiple state-of-the-art lossless and lossy compressors, resulting in significantly improved compression ratios.
  We evaluated this framework against seven X-CT datasets and observed a relative compression ratio improvement of 12.34× compared to the standard compression.
\end{abstract}
\maketitle
\input{01_introduction}
\input{02_background}
\input{03_relatedwork}
\input{04_approach}
\input{05_experiment}

\input{06_conclusion}
\input{07_ack}
\bibliographystyle{ACM-Reference-Format}
\bibliography{references}
\end{document}

%% file: 01_introduction.tex
\section{Introduction}

Synchrotron radiation facilities such as ESRF \cite{esrf}, APS \cite{aps}, and SPring-8 \cite{spring8} specialize in the generation of high-intensity X-rays. These facilities provide advanced material analysis and imaging across various scientific disciplines. In recent years, detectors at these light source facilities have become increasingly efficient: they are now capable of generating data ranging from terabytes to petabytes per day. This exponential growth of data creates significant challenges in storage, processing, and analysis that demand highly efficient data management strategies. To effectively handle these datasets, distributed storage architectures with compression capabilities must be integrated into high performance computing (HPC) environments \cite{Ishikawa2014}.

X-ray computed tomography (X-CT) images created by these large radiation facilities are particularly important because of their applications in various scientific, industrial, and medical domains. High-resolution imaging enables detailed investigation of internal material properties and structures \cite{XCT}. 
The data generation rate in SPring-8 has increased significantly since the introduction of DIFRAS detectors\cite{Kameshima2022}. This exponential growth of data creates challenges in storage, processing, and analysis that demand highly efficient data management strategies.
Recent variants of DIFRAS detectors equipped with the IMX661 ($13.9k$ x $9.7k$ pixels), $21.8$ frames/sec @ $10$ bit depth generate data at a rate of \SI{10.4}{GB/s} maximum (\SI{899}{TB/day})\cite{sony_sensor}. Therefore, data compression is critical for efficient storage and data analysis.

Traditional data compression approaches typically apply a general-purpose compression algorithm directly to X-CT images. These techniques do not achieve optimal results because they do not account for the specialized nature of X-CT data, including its unique noise patterns, spatial correlations, and critical structural features. Moreover, these standard methods often struggle with the high dynamic range found in X-CT scans, where both high-density and low-density materials must be preserved with high fidelity.

Lossless compression strategies such as Zstd \cite{Collet2016}, Gzip \cite{GZIP}, Huffman coding \cite{Huffman1952}, Bzip2 \cite{Seward1996}, and Lz77 \cite{Ziv1977} prioritize data fidelity over compression ratio and rate, which limits their effectiveness for large image datasets. In contrast, lossy compression methods such as Sz3 \cite{liang2021sz3modularframeworkcomposing} and Zfp \cite{doi:10.1177/10943420241284023} achieve high data reduction while preserving key characteristics of the object, such as precision, by removing less significant information.

Our approach employs region-of-interest (ROI) recognition to automatically segment significant structural regions from background elements based on density distributions. 
Because processing whole volumetric scans is computationally expensive and storage intensive, this domain-specific methodology extracts only the object of interest, effectively eliminating non-valuable areas entirely from further processing. By focusing exclusively on these relevant regions, we optimize both the reconstruction and segmentation steps. This reduction in working dataset size delivers dual benefits: speeding up data-intensive operations while minimizing storage and transmission costs.

Building on this foundation, we introduce a specialized pre-processing algorithm that conditions the segmented X-CT data to better align with the capabilities of standard compressors. This conditioning step enables significantly higher compression ratios compared to applying compressors directly to raw X-CT data while maintaining essential data quality throughout the process.

The key contributions of this paper include:
\begin{itemize}
    \item The development of an adaptive thresholding and binarization framework for processing 2D X-CT images.
    \item The Implementation of a region/object extraction strategies to isolate diagnostically relevant areas from X-CT data.
    \item The integration of a region-of-interest recognition with error-bounded quantization to analyze the relationship between compression ratio and data preservation within specified error tolerances.
    \item The demonstration that pre-processing segmentation enhances data processing efficiency while reducing storage requirements for X-CT datasets.
    \item The evaluation of the approach across multiple datasets, comparing compression ratios with state-of-the-art compressors to validate performance improvements.
\end{itemize}

This paper is organized as follows. 
Section~\ref{Section:2} provides background on X-CT and reviews the fundamental techniques. 
Section~\ref{Section:3} discusses related work in X-CT data compression and ROI extraction. 
Section~\ref{Section:4} presents our object extraction strategy, while Section~\ref{Section:5} evaluates our approach across multiple datasets, analyzing the compression ratio, processing time, and reconstruction quality. 
Finally, we conclude with a summary of the findings and directions for future research.

%% file: 02_background.tex
\section{Background}
\label{Section:2}
\begin{figure}[h]  
    \centering
    \includegraphics[width=0.5\textwidth]{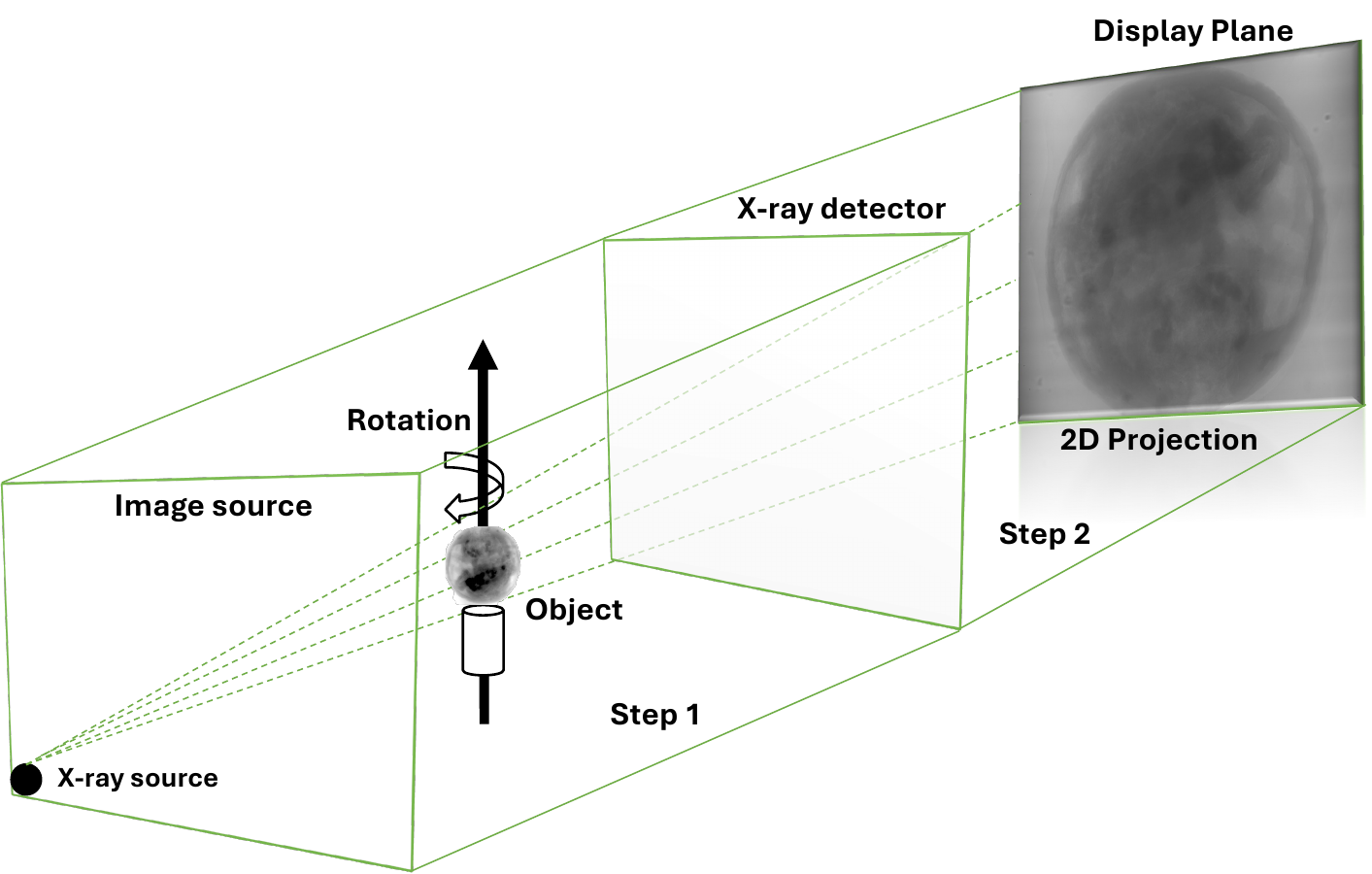}  
    \caption{Projection of X-CT Object Capturing Structural and Density}
    \label{fig:2D}
\end{figure}

X-ray Computed Tomography (X-CT) is an imaging technique that provides a high-resolution 3D representation of an object's internal structure. X-CT acquires numerous 2D X-ray projections from different angles, as illustrated in Figure~\ref{fig:2D}. During the imaging process, the X-ray source emits radiation toward the object, which is positioned between the source and the detector. The object undergoes stepwise rotation to capture multiple projections from different angles, as shown in Step 1 of Figure~\ref{fig:2D}. These projections are then recorded by the detector and processed to generate 2D projection images (Step 2).

The analysis of X-CT scans involves multiple phases, such as pre-processing, shape identification, feature extraction based on pixel intensity and structural patterns, classification, and post-processing. 
Feature extraction separates structures or materials of interest from the background. 
In the context of large-scale scientific imaging, effective segmentation techniques enable the isolation of critical regions by filtering out less relevant information.
This technique reduces computational overhead, facilitates more accurate measurements, and simplifies subsequent data handling, storage, and analysis tasks. 

X-CT data offer significant compression opportunities because most scan content is not critical: in typical scans, only the small region of interest requires high resolution detail, while large areas outside contain minimal valuable information. 
By recognizing this pattern, we can apply selective compression that preserves quality in important areas while aggressively reducing data in noncritical regions. 

%% file: 03_relatedwork.tex
\section{Related Work}
\label{Section:3}

Several studies have explored region-of-interest-based techniques in X-CT, addressing challenges related to data efficiency, reconstruction, and segmentation. Our work addresses ROI-based compression, which requires both effective ROI identification and efficient compression methods.

Existing methods often depend on lossless compression, which ensures exact data reconstruction but struggles to achieve high compression ratios, particularly for large-scale structured datasets. As a result, lossy compression approaches have emerged as more suitable alternatives to handle high-dimensional scientific data. Specifically, the lossy compressors Sz3~\cite{liang2021sz3modularframeworkcomposing} and Zfp~\cite{doi:10.1177/10943420241284023} are used to compress floating-point data. These compressors are designed to handle scientific datasets with precision control, adaptive compression, and high-throughput performance. Sz3 utilizes predictive coding and quantization to reduce redundancy with controlled accuracy loss, while Zfp uses blockwise transform coding to enable compression with fixed precision constraints.

ROI-based approaches have been successfully applied to various X-CT processing tasks, demonstrating the value of region-specific prioritization. In reconstruction contexts, X-CT imaging faces computational challenges including high radiation exposure and expensive forward projection steps. Huck et al.~\cite{Huck2021} explored ROI-based X-CT using a dynamic beam attenuator (DBA) to minimize radiation exposure while maintaining diagnostic accuracy. McCann et al.~\cite{8363524} proposed an ROI reconstruction method for parallel X-CT that eliminates the costly forward projection step through digital filtering implementations.

For ROI identification, effective segmentation methods are essential as pre-processing steps. Zheng~\cite{Zheng2024} proposed a lung CT segmentation algorithm that combines thresholding and gradient-based methods, utilizing time-evolutionary feature extraction to improve segmentation accuracy. Muthukrishnan~\cite{MUTHUKRISHNAN2024100215} proposed a topological derivative approach for CT and MRI segmentation in head and brain imaging. However, Müller et al.~\cite{muller2022guideline} identified critical evaluation challenges, finding that common segmentation metrics produce misleading results due to the class imbalance between small ROIs and large background areas.

Recent advances in ROI-based compression have shown promising results, though primarily in medical imaging applications. The Swin Transformer-based ROI model~\cite{roi_compression} demonstrated how ROI mask frameworks can enhance key region quality. Wasson~\cite{kaur2015roi} developed a hybrid compression technique for telemedicine that applies lossy fractal compression to non-ROI regions while preserving ROI areas with lossless compression. However, their approach lacks integration with high-performance computing frameworks and faces limitations in scalability for large-scale X-CT datasets.

ROI-based HPC compression has not yet been fully augmented with these advanced techniques, emphasizing the need for further research to integrate ROI-aware prioritization methods with high-performance scientific data compression. Current scientific compression frameworks such as Sz3 and Zfp apply uniform compression across entire datasets~\cite{liang2021sz3modularframeworkcomposing, doi:10.1177/10943420241284023}, without considering the varying scientific importance of different regions within the data. Although ROI-driven models have been widely applied to images, particularly in medical imaging~\cite{kaur2015roi} and remote sensing applications~\cite{roi_compression}, there is a significant gap in integrating ROI-based prioritization with error-bounded compression in large-scale scientific data. Current ROI methods typically operate at the pixel level in image compression, but scientific datasets often require region-specific precision control across entire data volumes. An improvement lies in developing hybrid compression methods that apply ROI prioritization to structured datasets, allowing Sz3 and Zfp to maximize efficiency without compromising scientific accuracy. Bridging this gap would improve data reduction strategies for high-resolution imaging, large-scale simulations, and scientific computing applications requiring precise information retention.

%% file: 04_approach.tex
\section {Proposed Approach}
\label{Section:4}
\begin{figure*}[h]  
    \centering
    \includegraphics[width=0.9\textwidth]{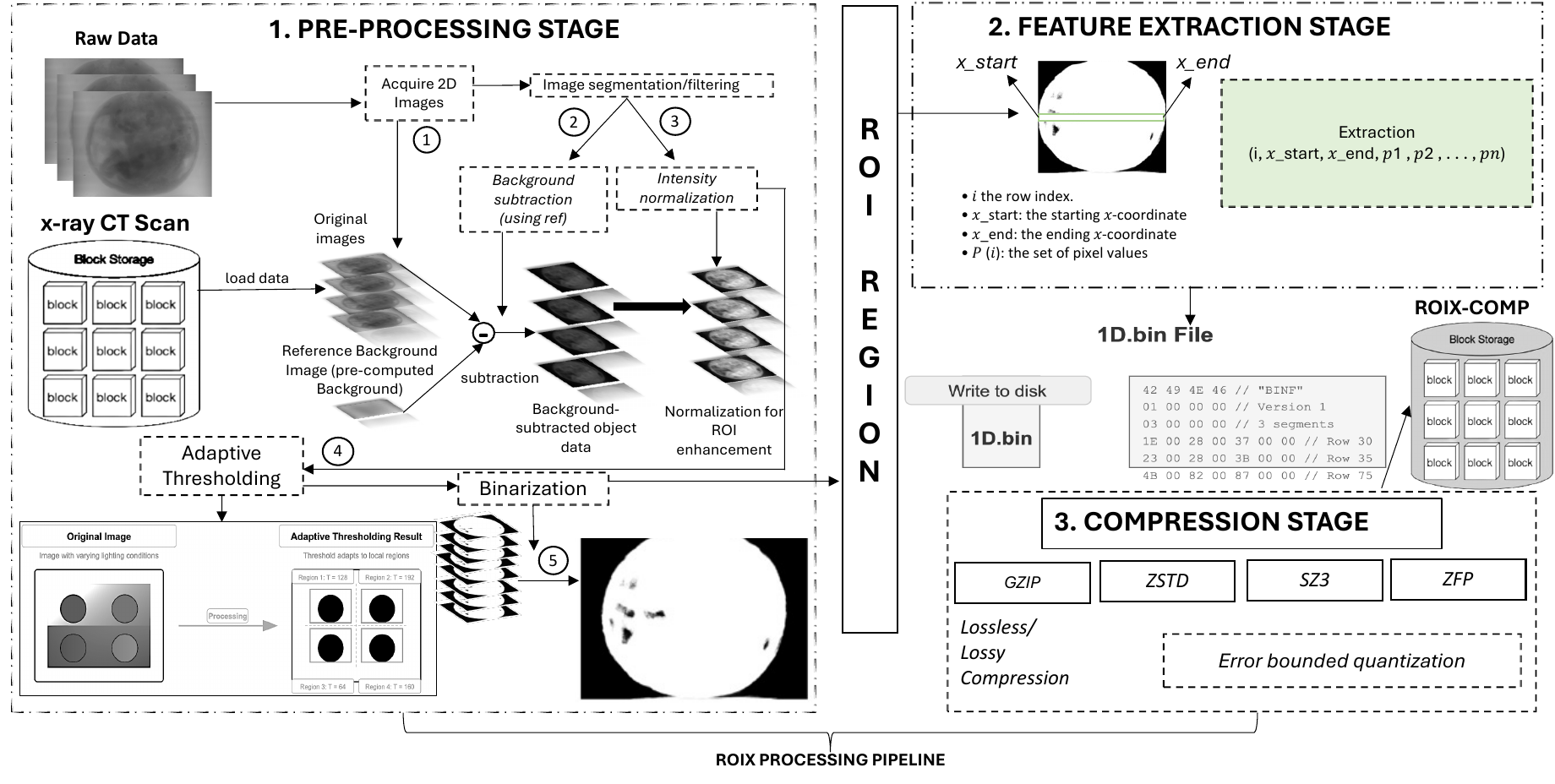}  
    \caption{Adaptive ROIX Compression Framework: Systematic Workflow}
    \label{fig:workflow1}
\end{figure*}

We propose a 2D X-CT extraction framework, illustrated in Figure~\ref{fig:workflow1}, which consists of several key steps: 1) Pre-processing (cf. Section \ref{subsection:preprocessing}), 2) Feature extraction (cf. Section \ref{subsection:extraction}), and 3) HPC-based compression (cf. Section \ref{subsection:compression}).

\subsection{Pre-processing Stage}
\label{subsection:preprocessing}

Before applying ROI-based compression, the raw X-CT data must be prepared to facilitate accurate region identification and consistent processing across different datasets. The pre-processing consists of four phases: 1) a substraction phase where the background is removed based on pixel intensity, 2) a normalization phase to avoid inconsistent analysis due to variations in imaging conditions, 3) an adaptive thresholding to automatically determine optimal threshold values based on local image characteristics, and 4) a binary masking to create precise region-of-interest boundaries for subsequent analysis.





\subsubsection{Background subtraction:}
X-CT scans exhibit a characteristic structure with distinct regions of varying scientific importance. We define the **foreground** as the central region containing the object or features of scientific interest, which typically corresponds to the specimen, material, or anatomical structure being analyzed in the middle of the X-CT scan. In contrast, we define the **background** as the surrounding areas that contain no valuable information for scientific analysis, such as empty space, mounting apparatus, or air gaps.

A key property of X-CT imaging is the clear intensity difference between the foreground and background regions as a result of varying material densities and X-ray absorption coefficients. The foreground objects typically exhibit higher intensity values compared to the relatively uniform low-intensity background areas. We leverage this intensity contrast to automatically distinguish between scientifically relevant and irrelevant regions.

To detect and remove the background based on this intensity difference, we employ a flexible approach with multiple options depending on the availability of data. Our solution primarily uses **background references** that are acquired during calibration scans without any sample present. When background references are unavailable, our method **estimates a static background** from the scene by analyzing regions where no sample is present and extrapolating these values across the image. 

This approach provides flexibility by utilizing highly accurate background references when available, while still functioning effectively through background estimation when necessary. Once the background is established, our method removes it by pixel-by-pixel subtraction from the original image.

Given an input image \( I(x, y) \) (where \( x \) and \( y \) represent spatial coordinates in the image) and a background reference \( B(x, y) \), we compute the output image \( M(x, y) \) with the following formula:

\[
M(x, y) = I(x, y) - B(x, y)
\]

By removing the background, we preserve all essential information while enabling targeted compression that focuses computational resources on scientifically significant features.

\subsubsection{Intensity normalization:}
The conditions of acquisition of X-CT images (Section~\ref{Section:2}) may vary during the procedure. These variations in scanning parameters, X-ray source intensity, or detector sensitivity can result in inconsistent images within the dataset. This variability can negatively impact subsequent processing steps, particularly segmentation and feature extraction. To reduce this variability, we normalize all image pixels to fit within an 8-bit range (0-255). We selected this range because it provides sufficient dynamic range for distinguishing relevant material densities while standardizing the input for our optimized compression algorithms. This normalization ensures consistent processing regardless of the original acquisition conditions.

Given an input grayscale image \( I(x, y) \) with pixel values varying from \( 0 \) to \( I_{\max} \), the normalization process is computed as:

\begin{equation}
I_n(x, y) = \frac{I(x, y)}{I_{\max}} \times 255
\end{equation}

where $I_{\max}$ is the maximum pixel intensity in the image.

Intensity normalization is applied to the background-subtracted object data (not the raw projections) to enhance the visibility of thin layers and low-intensity features that are difficult to detect at certain projection angles. This normalization aids ROI extraction while preserving the physical intensity relationships in the original data.

\subsubsection{Adaptive Thresholding:}
After intensity normalization of the X-CT images, we applied a thresholding and binarization procedure to separate the object from its background. 
Thresholding works by setting a specific intensity value as a cut-off point: pixels above this value are considered part of the object, while pixels below are classified as background. 

We specifically used the multi-Otsu adaptive thresholding technique \cite{Liao2001}, which extends the Otsu method \cite{Otsu1979} to compute multiple threshold values and segment images into classes based on intensity distribution. Unlike conventional fixed-threshold approaches, this method dynamically determines optimal threshold classes, making it effective for X-CT images with varying intensity levels.

\subsubsection{Binarization:}
Once the threshold is found, we proceed with the binarization.
Binarization creates a representation, called a binary mask M(x,y), where the object pixels have a value of 1 and the background pixels have a value of 0 isolating the object of interest:
\begin{equation}
M(x, y) = \begin{cases}
1, & \text{if pixel belongs to segmented ROI} \\
0, & \text{otherwise}
\end{cases}
\end{equation}

This step eliminates noise and irrelevant pixels, ensuring that subsequent analysis focuses exclusively on the target structures or features of interest.

\subsection{Feature Extraction}
\label{subsection:extraction}
To extract a feature, we first need to determine its contour from the created binary mask.
\subsubsection{Contour detection:} 
We use OpenCV's \texttt{findContours()} function with \texttt{RETR\_EXTERNAL} mode to detect only outer boundaries and \texttt{CHAIN\_APPROX\_SIMPLE} for efficient coordinate storage. From all detected contours, we select the largest by area to focus on the main object.

\begin{figure*}[h]  
    \centering
    \includegraphics[width=0.7\textwidth]{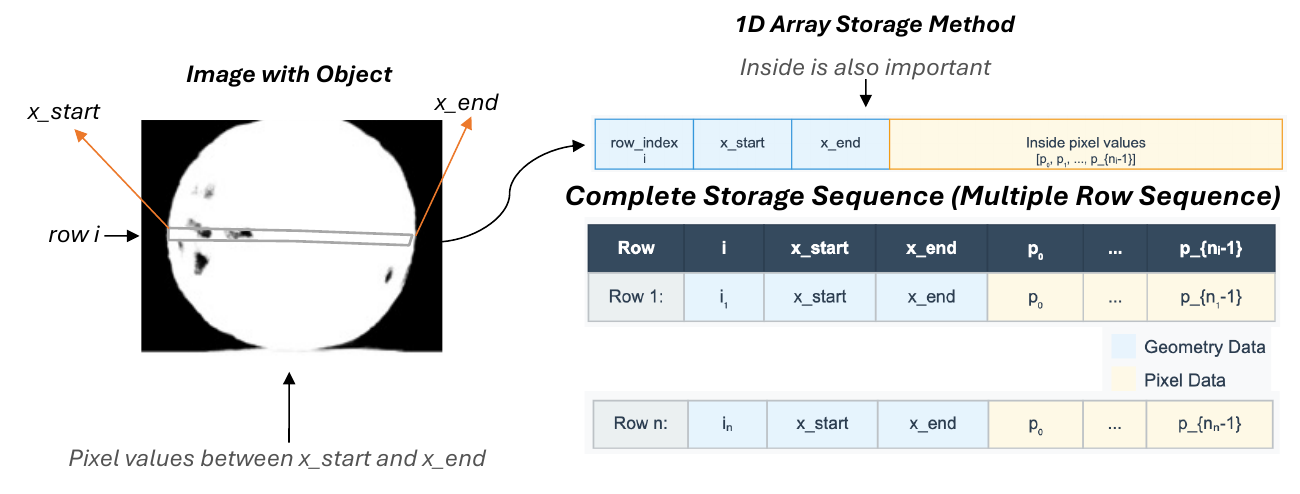}  
    \caption{ROI Extraction Pipeline: From Data to Targeted Feature Isolation}
    \label{fig:storage method}
\end{figure*}






Once ROI boundaries are identified, we convert the contour information into a row-based representation for efficient data organization. For each row that intersects with the ROI, we determine the leftmost and rightmost pixel coordinates that fall within the object boundaries. This creates a compact representation that separates structural information from pixel content.

\subsubsection{Feature extraction:} 
As illustrated in Figure \ref{fig:storage method}, for each row index $i$ that contains part of the object, we extract and store the following:

\textbf{Geometry Data Structure}
\begin{equation}
{G(i) = [i, x_{\text{start}}, x_{\text{end}}]}
\end{equation}


\textbf{Pixel Data Structure}
\begin{equation}
P(i) = [p_0, p_1, \ldots, p_{n_i-1}]
\end{equation}

Here we define:
\begin{itemize}
    \item $i$: the row index
    \item $x_{\text{start}}$: the starting $x$-coordinate of the object in row $i$
    \item $x_{\text{end}}$: the ending $x$-coordinate of the object in row $i$
    \item $p_j$: the pixel value at position $x_{\text{start}} + j$ in row $i$
    \item $n_i: x_{\text{end}} - x_{\text{start}}$: the number of pixels in row $i$
\end{itemize}

The complete ROI representation consists of:
\begin{itemize}
    \item \textbf{Geometry section} $\mathcal{G} = \{G(i_1), G(i_2), \ldots, G(i_m)\}$ containing spatial indexing data
    \item \textbf{Pixel section} $\mathcal{P} = \{P(i_1), P(i_2), \ldots, P(i_m)\}$ containing intensity values
\end{itemize}


This approach separates spatial indexing information from pixel content. We preserve the indexing data exactly to maintain accurate reconstruction coordinates. The intensity values of the pixels can be compressed with controlled quality loss. In this way, we maintain precise ROI positioning while achieving file size reduction.

\subsubsection{Absolute Error-Bounded Quantization}


After separating ROI data into indexing and pixel parts, we compress the pixel values to reduce storage space while keeping the indexing data exact. We use absolute error-bounded quantization because it controls quality loss by ensuring compressed pixel values differ from originals by no more than a set error limit. This method provides three key benefits: significant reduction in file size and maintains scientific accuracy. Unlike general compressors that treat all data uniformly, our approach targets ROI-specific compression, preserving structural information while compressing pixels with controlled precision for scientific applications.

\begin{description}
\item[\textbf{Input:}] Pixel sequence $D[0], D[1], ..., D[m-1]$ (where $m$ is the sequence length), positive absolute error tolerance $E_{\text{abs}}$
\item[\textbf{Output:}] Quantized sequence $Q[0], Q[1], ..., Q[m-1]$ where $|Q[i] - D[i]| \leq E_{\text{abs}}$
\end{description}

\textbf{Step 1: Calculate Error Bounds}  
For each pixel, define its acceptable range:
\begin{align}
\text{upper}[i] &= D[i] + E_{\text{abs}} \\
\text{lower}[i] &= D[i] - E_{\text{abs}}
\end{align}

\textbf{Step 2: Initialize}  
\begin{equation}
u = +\infty, \quad l = -\infty, \quad \text{head} = 0
\end{equation}
where $[l, u]$ tracks the intersection of acceptable ranges for the current group.  
\textit{Note: The midpoint is only computed after the actual data have updated the group range $[l, u]$, avoiding the undefined behavior of the initial values $\pm\infty$.}

\textbf{Step 3: For each pixel $i = 0$ to $m-1$:}
\begin{enumerate}
    \item Compute intersection with the 
    current group: \begin{align}
    l_{\text{new}} &= \max(l, \text{lower}[i]) \\
    u_{\text{new}} &= \min(u, \text{upper}[i])
    \end{align}
    
    \item Check if ranges overlap:
    \begin{itemize}
        \item \textbf{If} $u_{\text{new}} < l_{\text{new}}$ (no overlap):
        \begin{align}
        Q[j] &= \left\lfloor \frac{u + l}{2} \right\rfloor \quad \forall j \in [\text{head}, i-1] \\
        l &= \text{lower}[i] \\
        u &= \text{upper}[i] \\
        \text{head} &= i
        \end{align}
        
        \item \textbf{Else} (ranges overlap):
        \begin{align}
        l &= l_{\text{new}} \\
        u &= u_{\text{new}}
        \end{align}
    \end{itemize}
\end{enumerate}

\textbf{Step 4: Final Group}  
\begin{equation}
Q[j] = \left\lfloor \frac{u + l}{2} \right\rfloor \quad \forall j \in [\text{head}, m-1]
\end{equation}
\textit{The midpoint of the acceptable range is chosen to minimize the absolute worst-case  error across all pixels in the group.}

\textbf{Step 5: Clip to Valid Range}  
\begin{equation}
Q = \text{clip}(Q, 0, 255)
\end{equation}

\textbf{Usage:} This absolute-error bounded quantization method serves as a preparation step for general purpose compressors (Gzip, Zstd) to achieve error-bounded compression. Specialized error-bounded quantization compressors (Sz3, Zfp) bypass this step because they incorporate their own error-control mechanisms.

\subsection{Integrating HPC-based compression}
\label{subsection:compression}


At this point, only the regions of interest from the X-CT image have been preserved and prepared for compression. ROIX-Sz3 and ROIX-Zfp utilize the compressors' built-in error control mechanisms: Sz3 employs prediction-based compression with adjustable parameters, while Zfp uses block-based compression with fixed-rate or fixed-precision options. Both specialized compressors can adapt their strategies to preserve critical features in important regions. This combination of ROIX's targeted data selection with both general-purpose and domain-specific compression techniques enables comprehensive evaluation across different scientific data types and application requirements.

Our compression methodology begins with adaptive thresholding to identify regions of interest. After object detection, we extract their geometric data and convert it into a compact one-dimensional array stored as a binary file (.bin). This 1D representation preserves critical structural elements: row indices, x-coordinate boundaries, pixel values, and intensities. The compressed format consists of this binary file that contains the quantized 1D array based on our error-bounded constraints. We then apply either lossless methods (Gzip, Zstd) or lossy scientific compressors (Sz3, Zfp) to this representation.

During decompression, our algorithm reconstructs the image through a systematic reverse process. Each row is precisely restored using the preserved metadata captured during compression, specifically the row indices, x-coordinate boundaries, pixel values, and intensities. This comprehensive approach ensures that the spatial relationships and shape of the object are accurately reproduced. The reconstructed object is then combined with the precomputed background to recreate the original image.

%% file: 05_experiment.tex
\section{Experimental Evaluation}
\label{Section:5}

We conduct our evaluations across six key components: (1) ROI evaluation measuring spatial reduction efficiency, (2) **ROI evaluation** using standard metrics (DSC, IoU, sensitivity, specificity, accuracy, Kappa coefficient, AUC, and Average Hausdorff Distance) to validate extraction accuracy, (3) compression ratio analysis comparing lossless compression with relative improvements to demonstrate our method's effectiveness, followed by error-bounded quantization analysis across different error bounds to achieve higher compression ratios, (4) compression time evaluation, (5) decompression time assessment for reconstruction efficiency, and (6) quality assessment using Structural Similarity Index (SSIM) to validate preservation of meaningful content. This multifaceted evaluation demonstrates both the effectiveness of our ROI extraction methodology and the practical benefits achieved through integrated compression optimization.

\textbf{Hardware Configurations:} Our experiments were conducted on an x86-64 system (SYS-4029GP-TRT2) equipped with Intel(R) Xeon(R) Gold 5215 CPU @ 2.50GHz and an NVIDIA A100 GPU with 80 GB of memory.

\textbf{Software Environment:} Our implementation was developed using a Python environment. For the Sz3 and Zfp compression algorithms, we utilized libpressio integration via spack.

\textbf{Dataset:} We tested our approach against several X-CT datasets; their details are provided in Table~\ref{tab:X-CTData}.

\begin{table}[h]
    \centering
    \footnotesize  
    \renewcommand{\arraystretch}{0.8}
    \resizebox{\columnwidth}{!}{
\begin{tabular}{|c|c|c|c|}
\hline
\textbf{Dataset name} & \textbf{Resolution} & \textbf{Number of images} & \textbf{Total Size (GB)} \\ \hline
\textbf{Wood} \cite{WoodDataset2022}     & 740x300   & 900  & 0.4  \\ \hline
\textbf{Fossil embryos of cnidarians} \cite{XrayCTDataset2022}   & 4090x3008 & 1804 & 42   \\ \hline
\textbf{Ryugu} \cite{RyuguDataset2023}    & 1207x659  & 903  & 1.4  \\ \hline
\textbf{Chicken} \cite{meaney_2022_6990764}  & 2240x2368 & 721  & 3.83 \\ \hline
\textbf{Walnut} \cite{meaney_2022_6990764}   & 2240x2368 & 721  & 7.66 \\ \hline
\textbf{Pine cone} \cite{meaney_2022_6990764} & 2240x2368 & 721  & 7.66 \\ \hline
\textbf{Seashell} \cite{meaney_2022_6990764} & 2240x2368 & 721  & 7.66 \\ \hline
\end{tabular}
}
    \caption{X-CT Dataset Specification}
    \label{tab:X-CTData}
\end{table}

\subsection{ROI detection evaluation}

The evaluation demonstrates the framework's ability to accurately identify and extract object structures while eliminating background areas. 

\begin{figure*}[htbp]
\centering
\begin{tikzpicture}
\begin{axis}[
    ybar,
    bar width=0.5cm,
    width=0.9\textwidth,
    height=6cm,
    enlarge x limits=0.15,
    ylabel={Spatial Reduction Factor (×)},
    xlabel={Dataset},
    symbolic x coords={Wood, Fossile, Ryugu, Chicken, Walnut, Pine Cone, Seashell},
    xtick=data,
    x tick label style={rotate=45, anchor=east},
    nodes near coords,
    nodes near coords align={vertical},
    nodes near coords style={font=\footnotesize},
    ymin=0,
    ymax=10,
    grid=major,
    grid style={dashed, gray!30},
]
\addplot[
    fill=blue!70!black,
    draw=black,
    line width=0.5pt
] coordinates {
    (Wood, 2.45)
    (Fossile, 1.51)
    (Ryugu, 8.49)
    (Chicken, 4.87)
    (Walnut, 2.62)
    (Pine Cone, 4.66)
    (Seashell, 3.81)
};
\end{axis}
\end{tikzpicture}
\caption{Spatial reduction achieved through ROI extraction across different datasets. The spatial reduction factor indicates how many times smaller the ROI is compared to the original image.}
\label{fig:spatial_reduction}
\end{figure*}
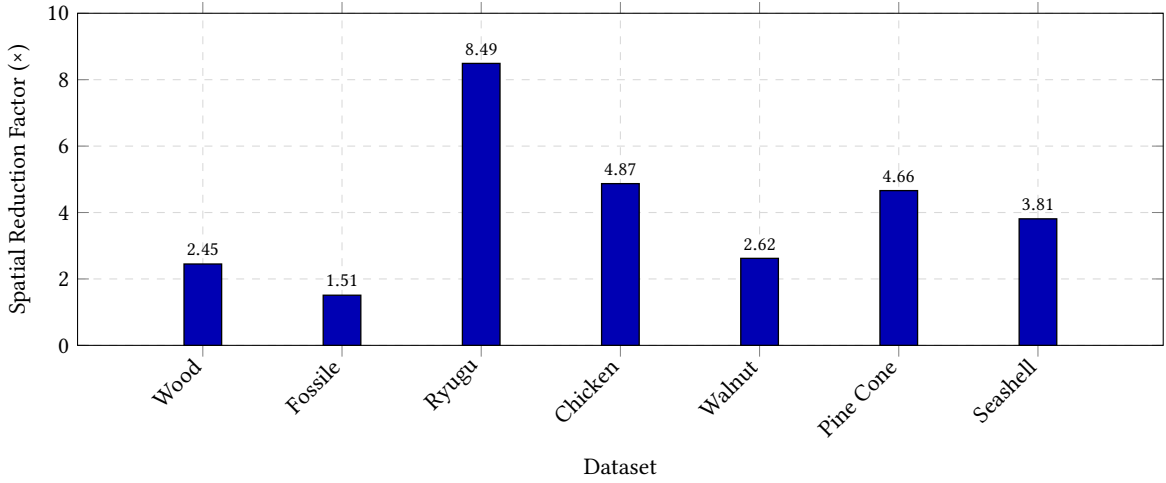

Figure~\ref{fig:spatial_reduction} demonstrates the effectiveness of the proposed ROI extraction methodology in different datasets, highlight significant variations in spatial reduction factors that reflect the inherent characteristics of different image types. The results show that the Ryugu dataset achieved the highest spatial reduction factor of $8.49\times$. In contrast, the Fossile dataset exhibited the lowest spatial reduction factor of $1.51\times$, implying that meaningful features are more widely distributed throughout the image, requiring a larger ROI that encompasses approximately $66\%$ of the original image area. The remaining datasets (Chicken: $4.87\times$, Pine cone: $4.66\times$, Seashell: $3.81\times$, Walnut: $2.62\times$, and Wood: $2.45\times$) demonstrate intermediate spatial reduction factors, ranging from $2.45\times$ to $4.87\times$. These varying reduction factors highlight the adaptability of the ROI extraction algorithm to different types of image content, with the average spatial reduction factor of $4.06\times$ in all datasets indicating substantial potential for storage and bandwidth optimization in practical applications. The significant variation in spatial reduction factors underscores the importance of content-adaptive ROI identification methods that can automatically adjust to the spatial distribution characteristics of different image types.

\subsection{ROI quality evaluation}
To comprehensively evaluate our ROI detection method, we used multiple standard metrics that measure different aspects of segmentation performance. The Dice Similarity Coefficient (DSC) and Intersection over Union (IoU) assess the overlap accuracy between our detected regions and ground truth object. Sensitivity measures how well our method identifies all relevant pixels, while specificity evaluates its ability to correctly exclude background areas. 
Accuracy provides general correctness, and the Kappa coefficient accounts for chance agreement by measuring performance beyond random classification.
The Area Under Curve (AUC) summarizes performance across different threshold values, and the Average Hausdorff Distance (AHD) measures boundary precision by calculating the maximum distance between predicted and actual boundaries. Together, these metrics provide a complete picture of our method's detection accuracy, boundary precision, and reliability across different object types and imaging conditions.

The evaluation metrics are mathematically defined as follow:

\begin{equation}
DSC = \frac{2|A \cap B|}{|A| + |B|}
\end{equation}

\begin{equation}
IoU = \frac{|A \cap B|}{|A \cup B|}
\end{equation}

\begin{equation}
Sensitivity = \frac{TP}{TP + FN}
\end{equation}

\begin{equation}
Specificity = \frac{TN}{TN + FP}
\end{equation}

\begin{equation}
Accuracy = \frac{TP + TN}{TP + TN + FP + FN}
\end{equation}

\begin{equation}
f_c = \frac{(TN + FN)(TN + FP) + (FP + TP)(FN + TP)}{TP + TN + FN + FP}
\end{equation}

\begin{equation}
Kappa = \frac{(TP + TN) - f_c}{(TP + TN + FN + FP) - f_c}
\end{equation}

\begin{equation}
AUC = 1 - \frac{1}{2}\left(\frac{FP}{FP + TN} + \frac{FN}{FN + TP}\right)
\end{equation}

\begin{equation}
AHD = \max\left(\frac{1}{|A|}\sum_{a \in A}\min_{b \in B}d(a,b), \frac{1}{|B|}\sum_{b \in B}\min_{a \in A}d(a,b)\right)
\end{equation}

where $A$ represents the predicted ROI, $B$ represents the ground truth ROI, $TP$ is true positive, $TN$ is true negative, $FP$ is false positive, $FN$ is false negative, $f_c$ is the expected agreement by chance, and $d(a,b)$ is the Euclidean distance between the points $a$ and $b$.

\begin{figure*}[h]  
    \centering
    \includegraphics[width=1\textwidth]{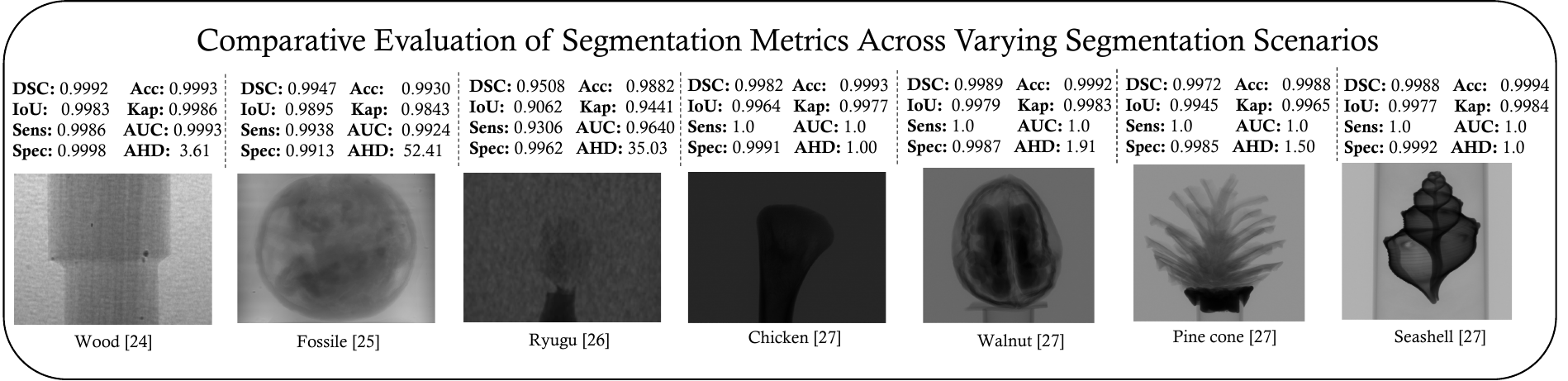}  
    \caption{Evaluation metrics measuring different aspects of segmentation performance: overlap accuracy (DSC, IoU), detection capability (Sensitivity, Specificity), overall correctness (Accuracy), and boundary precision (AHD).}
    \label{fig:Dd}
\end{figure*}

 This paper evaluates the effectiveness of the proposed method using the Dice Similarity Coefficient (DSC), sensitivity, and other metrics\cite{muller2022guideline} based on image segmentation. Therefore, we focus on showing how our method performs on various metrics, avoiding common evaluation mistakes in this field. Our evaluation of seven test cases demonstrates that our segmentation method achieves strong performance across different scenarios. The Dice coefficient (DSC) ranged from 0.9508 to 0.9992, with six out of seven cases exceeding 0.99, indicating excellent segmentation quality. Case 3 showed the most interesting result: despite having the lowest DSC (0.9508), it still maintained high accuracy (0.9882), demonstrating how accuracy can be misleading by showing only a 3.7 percent drop while actual segmentation quality dropped nearly 5 percent. This confirms that accuracy inflates performance scores in X-CT imaging due to large background areas. Cases 4-7 achieved perfect sensitivity (1.0), showing complete detection of all ROI pixels, while Cases 1-3 had slightly lower sensitivity (0.9306-0.9986). Especially, case 2 revealed a critical finding: despite the excellent DSC (0.9947), it had the highest Average Hausdorff Distance (52.41 pixels), indicating poor boundary precision that overlap metrics alone cannot detect. In contrast, cases 4-7 showed both high DSC (>0.997) and low AHD (1.0-1.91), representing optimal performance. These results \ref{fig:Dd} confirm our method performs excellently overall, with only Case 3 showing moderate performance and Case 2 highlighting the importance of using boundary-sensitive metrics alongside overlap-based measures.


\subsection{Compression Ratio evaluation}
We evaluated the compression performance of different compressors applied to multiple X-CT datasets without error bound quantization (lossless). 

\begin{figure*}[h]  
    \centering
    \includegraphics[width=0.8\textwidth, height=0.3\textheight]{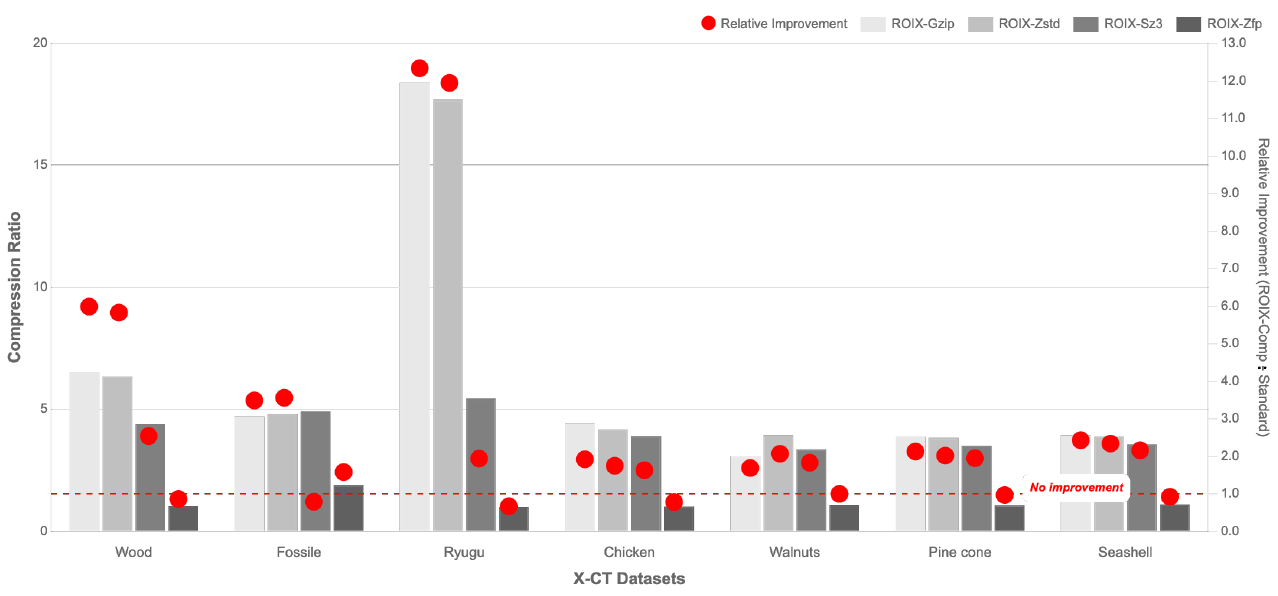}
    \caption{Comparative Analysis of Compression Ratios Across X-CT Datasets with relative improvements} 
    \label{fig:crwithimprovement}
\end{figure*}

We have evaluated the compression ratio (CR) illustrated in Figure \ref{fig:crwithimprovement} and compared the compression ratio performance of ROIX-Gzip, ROIX-Zstd, ROIX-Sz3, and ROIX-Zfp across multiple X-CT datasets. Among the compressors tested, ROIX-Gzip and ROIX-Zstd consistently achieve higher compression ratios across most datasets. ROIX-Gzip reaches a compression ratio of 18.39 in the ryugu dataset \cite{RyuguDataset2023}. ROIX-Sz3 demonstrates moderate compression efficiency. The results indicate that ROIX-Gzip outperforms other methods when no quantization is applied. In terms of relative improvement (ROIX-Comp ÷ Standard), the ryugu dataset \cite{RyuguDataset2023} shows the best performance, achieving improvement factors of 12.34× and 11.95× respectively. The wood dataset \cite{WoodDataset2022} also benefits from our approach, with 5.99× and 5.83× improvements using the same compression methods. Although most datasets show positive results for various compression algorithms, the fossil dataset \cite{XrayCTDataset2022} reveals an interesting pattern where ROIX-Sz3 performs worse than the baseline (0.79×), suggesting that our approach may not universally improve all compression techniques for all types of data. Generally, ROIX-Gzip and ROIX-Zstd consistently deliver the best performance across most datasets, while ROIX-Zfp shows minimal improvements or even slight degradation in some cases. 

\newcommand{\shadedcell}[2]{%
  \cellcolor{black!#1}%
  \ifnum#1>50 \color{white}\else\color{black}\fi%
  #2%
}

\begin{table}[htbp]
\centering
\caption{Compression ratios for different datasets and methods across various error bounds}
\begin{tabular}{@{}llrrrrr@{}}
\toprule
\textbf{Dataset} & \textbf{Method} & \multicolumn{5}{c}{\textbf{Error Bound}} \\
\cmidrule(l){3-7}
 &  & \textbf{1e-1} & \textbf{1e0} & \textbf{5e0} & \textbf{1e1} & \textbf{1.5e1} \\
\midrule

\multirow{4}{*}{Wood} & ROIX-Gzip & 
\shadedcell{14}{6.53} & \shadedcell{14}{6.60} & \shadedcell{23}{10.07} & \shadedcell{43}{17.46} & \shadedcell{67}{27.01} \\
& ROIX-Zstd & 
\shadedcell{14}{6.35} & \shadedcell{14}{6.36} & \shadedcell{21}{9.10} & \shadedcell{45}{18.29} & \shadedcell{74}{29.47} \\
& ROIX-Sz3 & 
\shadedcell{9}{4.40} & \shadedcell{18}{8.05} & \shadedcell{42}{17.22} & \shadedcell{62}{25.02} & \shadedcell{100}{39.56} \\
& ROIX-Zfp & 
\shadedcell{0}{1.03} & \shadedcell{5}{2.92} & \shadedcell{7}{3.57} & \shadedcell{8}{4.03} & \shadedcell{8}{4.03} \\
\midrule
\multirow{4}{*}{Fossil} & ROIX-Gzip & 
\shadedcell{4}{4.71} & \shadedcell{4}{4.85} & \shadedcell{10}{10.04} & \shadedcell{48}{39.90} & \shadedcell{100}{80.34} \\
& ROIX-Zstd & 
\shadedcell{4}{4.81} & \shadedcell{4}{4.87} & \shadedcell{9}{8.87} & \shadedcell{51}{42.19} & \shadedcell{99}{79.62} \\
& ROIX-Sz3 & 
\shadedcell{4}{4.91} & \shadedcell{8}{8.20} & \shadedcell{22}{19.23} & \shadedcell{45}{36.96} & \shadedcell{93}{74.86} \\
& ROIX-Zfp & 
\shadedcell{0}{1.90} & \shadedcell{0}{2.00} & \shadedcell{2}{3.57} & \shadedcell{0}{2.20} & \shadedcell{0}{2.20} \\
\midrule
\multirow{4}{*}{Ryugu} & ROIX-Gzip & 
\shadedcell{34}{18.39} & \shadedcell{34}{18.39} & \shadedcell{49}{26.31} & \shadedcell{76}{40.29} & \shadedcell{100}{52.56} \\
& ROIX-Zstd & 
\shadedcell{32}{17.69} & \shadedcell{32}{17.69} & \shadedcell{41}{22.27} & \shadedcell{71}{37.81} & \shadedcell{96}{50.53} \\
& ROIX-Sz3 & 
\shadedcell{9}{5.44} & \shadedcell{14}{8.40} & \shadedcell{30}{16.28} & \shadedcell{43}{23.01} & \shadedcell{67}{35.64} \\
& ROIX-Zfp & 
\shadedcell{0}{1.01} & \shadedcell{4}{3.14} & \shadedcell{6}{3.88} & \shadedcell{7}{4.42} & \shadedcell{7}{4.42} \\
\midrule
\multirow{4}{*}{Chicken} & ROIX-Gzip & 
\shadedcell{1}{4.42} & \shadedcell{3}{8.26} & \shadedcell{53}{122.09} & \shadedcell{79}{183.04} & \shadedcell{90}{208.20} \\
& ROIX-Zstd & 
\shadedcell{1}{4.17} & \shadedcell{2}{6.32} & \shadedcell{51}{118.40} & \shadedcell{85}{197.25} & \shadedcell{100}{230.81} \\
& ROIX-Sz3 & 
\shadedcell{1}{3.91} & \shadedcell{3}{7.50} & \shadedcell{7}{16.24} & \shadedcell{12}{28.66} & \shadedcell{20}{46.35} \\
& ROIX-Zfp & 
\shadedcell{0}{1.02} & \shadedcell{1}{2.94} & \shadedcell{1}{3.61} & \shadedcell{1}{4.07} & \shadedcell{1}{4.07} \\
\midrule
\multirow{4}{*}{Walnuts} & ROIX-Gzip & 
\shadedcell{1}{3.08} & \shadedcell{5}{8.40} & \shadedcell{18}{27.82} & \shadedcell{34}{52.42} & \shadedcell{48}{73.74} \\
& ROIX-Zstd & 
\shadedcell{2}{3.94} & \shadedcell{4}{6.59} & \shadedcell{40}{62.40} & \shadedcell{70}{107.64} & \shadedcell{100}{153.57} \\
& ROIX-Sz3 & 
\shadedcell{1}{3.35} & \shadedcell{3}{6.28} & \shadedcell{7}{11.28} & \shadedcell{10}{16.44} & \shadedcell{13}{21.23} \\
& ROIX-Zfp & 
\shadedcell{0}{1.08} & \shadedcell{0}{1.48} & \shadedcell{1}{1.86} & \shadedcell{1}{2.60} & \shadedcell{2}{3.40} \\
\midrule

\multirow{4}{*}{Pine Cone} & ROIX-Gzip & 
\shadedcell{2}{3.87} & \shadedcell{5}{7.42} & \shadedcell{39}{54.83} & \shadedcell{72}{100.10} & \shadedcell{100}{138.36} \\

& ROIX-Zstd & 
\shadedcell{2}{3.83} & \shadedcell{4}{5.90} & \shadedcell{41}{56.94} & \shadedcell{70}{96.48} & \shadedcell{96}{132.31} \\

& ROIX-Sz3 & 
\shadedcell{2}{3.51} & \shadedcell{4}{6.22} & \shadedcell{7}{11.11} & \shadedcell{11}{16.15} & \shadedcell{14}{20.77} \\

& ROIX-Zfp & 
\shadedcell{0}{1.07} & \shadedcell{0}{1.62} & \shadedcell{1}{2.02} & \shadedcell{1}{2.70} & \shadedcell{2}{4.00} \\
\midrule

\multirow{4}{*}{Seashell} & ROIX-Gzip & 
\shadedcell{3}{3.94} & \shadedcell{7}{8.34} & \shadedcell{39}{43.43} & \shadedcell{68}{75.64} & \shadedcell{100}{109.97} \\

& ROIX-Zstd & 
\shadedcell{3}{3.88} & \shadedcell{5}{6.86} & \shadedcell{36}{40.68} & \shadedcell{62}{68.63} & \shadedcell{92}{101.80} \\

& ROIX-Sz3 & 
\shadedcell{2}{3.57} & \shadedcell{4}{5.63} & \shadedcell{8}{9.35} & \shadedcell{11}{12.94} & \shadedcell{14}{16.41} \\

& ROIX-Zfp & 
\shadedcell{0}{1.10} & \shadedcell{1}{1.65} & \shadedcell{1}{2.08} & \shadedcell{1}{2.70} & \shadedcell{3}{4.05} \\
\bottomrule
\end{tabular}
\label{tab:compression-per-dataset}
\end{table}

The results in Table~\ref{tab:compression-per-dataset} demonstrate the effect of the absolute error bound quantization ~\cite{9499386} on the compression ratio for different compressors using the X-CT datasets. As the absolute error bound increases, all compressors show an improvement in the compression ratio. We have evaluated multiple error bound levels, ranging from 1e-01 to 1.5e1, where 1e-01 represents minimal data reduction, and 1.5e1 represents a greater reduction with potentially greater information loss. 

Based on the compression ratio data visualized in our tables, several important patterns appear across the different datasets. When examining the compression performance through the per-dataset gradient coloring, we observe that ROIX-Zstd and ROIX-Gzip consistently achieve the highest compression ratios at higher error bounds (1e1 and 1.5e1), with the Chicken dataset showing exceptional results reaching up to 230.81× compression with ROIX-Zstd.
The wood\cite{WoodDataset2022} and ryugu\cite{RyuguDataset2023} datasets show more modest but still significant compression, while ROIX-Sz3 demonstrates more balanced performance across different error bounds. Most datasets exhibit a dramatic increase in compression ratio when moving from 1e0 to 5e0 error bounds, indicating a key threshold where lossy compression becomes significantly more effective. In contrast, ROIX-Zfp consistently shows the lowest compression performance across all datasets, rarely exceeding 4.5× even at the highest error bounds.
The per-dataset visualization approach effectively highlights that different datasets respond uniquely to compression methods, with objects like pine cone\cite{meaney_2022_6990764} and walnuts\cite{meaney_2022_6990764} showing exceptional compressibility with ROIX-Zstd (reaching 132-153× compression), while ROIX-Sz3 performs particularly well on the wood\cite{WoodDataset2022} dataset. This suggests that compression algorithm selection should be tailored to specific dataset characteristics for optimal results in practical applications.

\subsection{Compression time evaluation}
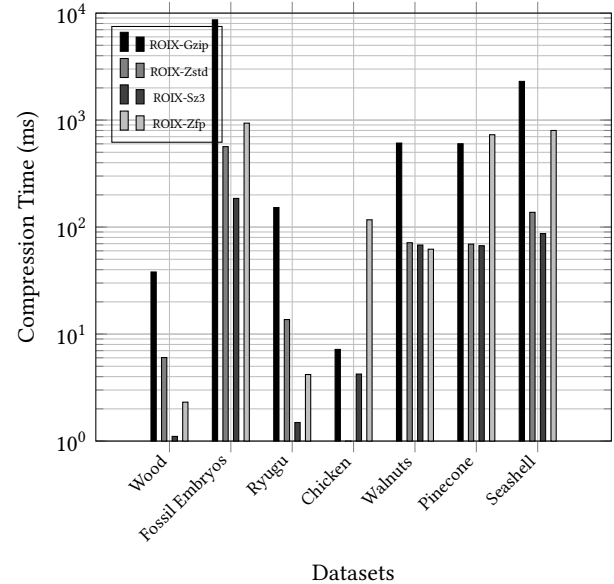
\begin{figure}[h]
    \centering
    \begin{tikzpicture}
        \begin{axis}[
            ybar,
            ymode=log, 
            log origin=infty, 
            symbolic x coords={Wood, Fossil Embryos, Ryugu, Chicken, Walnuts, Pinecone, Seashell},
            xtick=data,
            xticklabel style={font=\footnotesize, rotate=45, anchor=east},
            ymin=1, ymax=10000, 
            ylabel={Compression Time (ms)},
            xlabel={Datasets},
            bar width=2pt, 
            legend pos=north west,
            legend style={font=\tiny, fill=none},
            grid=both,
            enlarge x limits=0.2,
            ytick={1, 10, 100, 1000, 10000} 
        ]
        
        \addplot[fill=black] coordinates {
        (Wood,38.07) (Fossil Embryos,8656.19) (Ryugu,152.4) (Chicken,7.19) (Walnuts,610.25) (Pinecone,601.39) (Seashell,2301.29)
        };
        \addplot[fill=gray] coordinates {
        (Wood,6.03) (Fossil Embryos,565.14) (Ryugu,13.67) (Chicken,0.47) (Walnuts,71.45) (Pinecone,69.29) (Seashell,137.45)
        };
        \addplot[fill=darkgray] coordinates {
        (Wood,1.106) (Fossil Embryos,185.47) (Ryugu,1.491) (Chicken,4.23) (Walnuts,68) (Pinecone,67) (Seashell,87)
        };
        \addplot[fill=lightgray] coordinates {
        (Wood,2.31) (Fossil Embryos,936.86) (Ryugu,4.19) (Chicken,117) (Walnuts,62.14) (Pinecone,730) (Seashell,800)
        };
        
        \legend{ROIX-Gzip, ROIX-Zstd, ROIX-Sz3, ROIX-Zfp}
        
        \end{axis}
    \end{tikzpicture}
    \caption{Compression Time Performance Evaluation (Log Scale)}
    \label{fig:compression_time_log}
\end{figure}

The compression times presented in Figure~\ref{fig:compression_time_log} in seven X-CT datasets reveal significant performance differences between the four ROIX compression methods.

 ROIX-Zstd shows superior speed in most cases, particularly excelling with the chicken \cite{meaney_2022_6990764} dataset (0.47ms), while ROIX-Sz3 shows competitive performance, being fastest for the wood dataset (1.106ms). ROIX-Zfp performs well in certain datasets (wood \cite{WoodDataset2022}, ryugu \cite{RyuguDataset2023}), but exhibits notable slowdowns in others (pinecone \cite{meaney_2022_6990764}, seashell \cite{meaney_2022_6990764}), suggesting data-dependent characteristics. ROIX-Gzip consistently requires the longest processing times, particularly struggling with fossil \cite{XrayCTDataset2022} (8656.19ms) and seashell \cite{meaney_2022_6990764} (2301.29ms) datasets. The fossil dataset \cite{XrayCTDataset2022} presents the greatest challenge for all compressors, requiring significantly longer processing times for all methods. Performance differences span multiple orders of magnitude, and ROIX-Zstd and ROIX-Sz3 perform consistently faster than ROIX-Gzip, making them preferable choices for time-sensitive X-CT data processing applications.

\begin{figure}[htbp]
    \centering
    \begin{tikzpicture}
        \begin{axis}[
            ybar,
            ymode=log, 
            log origin=infty, 
            symbolic x coords={Wood, Fossil Embryos, Ryugu, Chicken, Walnuts, Pinecone, Seashell},
            xtick=data,
            xticklabel style={font=\footnotesize, rotate=45, anchor=east},
            ymin=0.5, ymax=2000, 
            ylabel={Decompression Time (ms)},
            xlabel={Datasets},
            bar width=2pt, 
            legend pos=north west,
            legend style={font=\tiny, fill=none},
            grid=both,
            enlarge x limits=0.2,
            ytick={1, 10, 100, 1000} 
        ]
        
        \addplot[fill=black] coordinates {(Wood,4.40) (Fossil Embryos,1046.08) (Ryugu,19.59) (Chicken,0.21) (Walnuts,103.34) (Pinecone,103.73) (Seashell,105.75)};
        \addplot[fill=gray] coordinates {(Wood,3.97) (Fossil Embryos,979.99) (Ryugu,11.27) (Chicken,0.2) (Walnuts,93.12) (Pinecone,144) (Seashell,100.31)};
        \addplot[fill=darkgray] coordinates {(Wood,0.58) (Fossil Embryos,122.96) (Ryugu,1.13) (Chicken,4.56) (Walnuts,98.2) (Pinecone,91) (Seashell,100.1)};
        \addplot[fill=lightgray] coordinates {(Wood,2.75) (Fossil Embryos,543.20) (Ryugu,4.77) (Chicken,62) (Walnuts,449) (Pinecone,256) (Seashell,359)};
        
        \legend{ROIX-Gzip, ROIX-Zstd, ROIX-Sz3, ROIX-Zfp}
        
        \end{axis}
    \end{tikzpicture}
    \caption{Decompression Time Comparison (Log Scale)}
    \label{fig:decompression_time_log}
\end{figure}
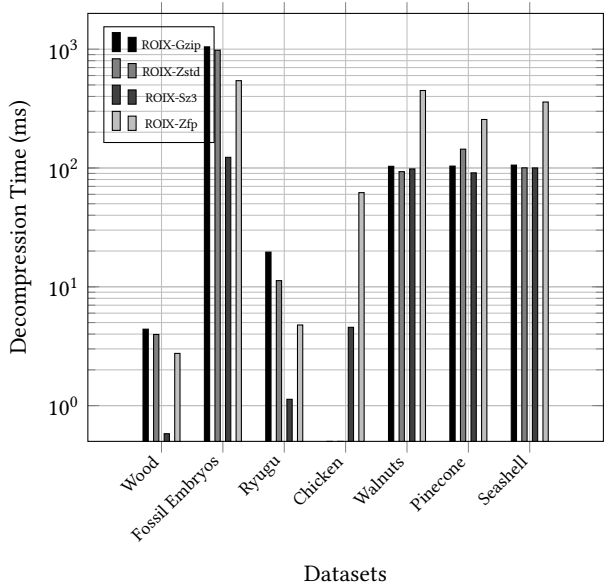

\subsection{Decompression time evaluation}
The decompression times shown in Figure~\ref{fig:decompression_time_log} follow a similar trend to the compression times. Comparison of decompression time demonstrates notable performance variations among the four ROIX methods. ROIX-Sz3 generally exhibits the fastest decompression speeds, particularly excelling with wood \cite{WoodDataset2022} (0.58ms) and ryugu \cite{RyuguDataset2023} (1.13ms) datasets, while showing consistent performance in most samples. ROIX-Gzip and ROIX-Zstd perform similarly to each other, with ROIX-Zstd showing slightly better results in some datasets, although both struggle with the more complex fossil dataset \cite{XrayCTDataset2022} (1046.08ms and 979.99ms, respectively). ROIX-Zfp consistently shows the slowest decompression times in most datasets, particularly with walnuts \cite{meaney_2022_6990764} (449ms), seashell \cite{meaney_2022_6990764} (359ms), and pinecone \cite{meaney_2022_6990764} (256ms). For time-critical applications requiring fast decompression, ROIX-Sz3 emerges as the optimal choice, with decompression performance advantages compared to other methods on certain datasets.

\subsection{Decompression quality evaluation}

Quantitative analysis was performed on reconstructed images derived from raw data. The evaluation focused specifically on targeted spatial features, enabling a comprehensive assessment of the selected structural elements.

\begin{figure}[h!]
    \includegraphics[width=0.51\textwidth]{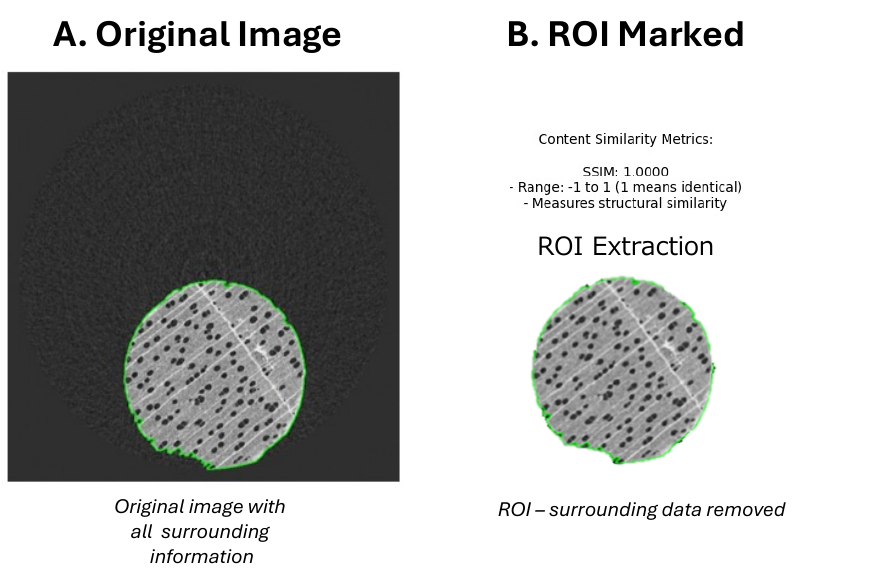}  
    \caption{Original vs ROI image analysis} 
    \label{fig:xct_processing}
\end{figure}

To verify the effectiveness of our object extraction process, we compared the original image with the reconstructed image after compression and decompression. We evaluated the quality and similarity between these images using the Structural Similarity Index (SSIM). Figure \ref{fig:xct_processing} highlights these metric scores and confirms that our method preserves objects with exceptional precision. For lossless compression configurations, the analysis demonstrates perfect structural preservation with an SSIM score of 1.0, indicating identical structural patterns between the original and reconstructed images. If we used lossy compression, it would likely affect the SSIM score, reducing it from the perfect 1.0 achieved with lossless methods. The degree of reduction would depend on the specific error-bounded settings applied during compression. These detailed measurements show that our extraction method preserves even the smallest details of the original image with high accuracy.

\section{Discussion}

Our ROI-based approach shows promising results but faces challenges with complex structures, particularly low contrast boundaries and irregular shapes. The exceptional compression performance of the Ryugu dataset demonstrates how image characteristics fundamentally impact efficiency: homogeneous backgrounds and well-defined boundaries achieve significantly better results than complex X-CT images with overlapping tissues and noise. Our analysis confirms that accuracy metrics can be misleading in X-CT imaging, as large background areas artificially inflate scores while masking segmentation quality degradation. Compression performance is strongly correlated with background homogeneity, noise levels, and ROI characteristics. Traditional compression algorithms show limited gains with ROI-extracted data, while HPC-based compressors performed particularly poorly. ZFP achieves only 3-4× compression even at high error bounds, significantly underperforming its typical behavior on scientific data. Possible factors include ZFP's block structure that interacts poorly with our ROI patch dimensions, 8-bit normalization reducing exploitable redundancy, or mismatch between normalized data characteristics and ZFP's floating-point design. More research is needed to fully characterize the underlying factors that contribute to these performance differences, which we identify as important future work. We recommend optimizing acquisition parameters to minimize background complexity and enhance tissue contrast. These findings highlight opportunities to develop image-characteristic-aware compression algorithms that automatically adapt to the properties of the dataset.

Although our results demonstrate effective ROI evaluation and compression performance, no direct comparison was performed with other state-of-the-art ROI methods on identical datasets. Fair comparison between different ROI approaches is complex due to varying optimization targets, preprocessing requirements, and dataset-specific parameter tuning needs. Our evaluation focused primarily on the integration of ROI detection with compression efficiency rather than comparative segmentation performance.

%% file: 06_conclusion.tex
\section{Conclusion} 

In this paper, we have proposed an ROI detection and extraction framework (ROIX-Comp) optimized specifically for compression efficiency while maintaining data quality. Our method applies intensity normalization and adaptive thresholding to improve object boundary detection and structure preservation in X-CT data. The framework supports both lossless and lossy compression methods, depending on the application requirements. The evaluation results demonstrate that compression performance is highly dependent on image characteristics and error-bounded quantization. Integrating ROI-based extraction with adaptive compression improves storage efficiency and computational performance for large-scale X-CT datasets, achieving up to 88\% data reduction while preserving valuable information.

Future work will include a comprehensive comparative evaluation against state-of-the-art ROI detection methods using identical datasets to establish relative performance benchmarks. We also plan to expand this framework to support diverse X-CT applications in materials science and X-CT imaging, integrate advanced deep learning-based segmentation models, and explore newer compression techniques tailored for scientific imaging to ensure optimal performance across various applications.

%% file: 07_ack.tex
\begin{acks}
This work has been supported by the COE research grant in computational science from Hyogo Prefecture and Kobe City through Foundation for Computational Science.
\\
This work ("AI for Science" supercomputing platform project) was supported by the RIKEN TRIP initiative (System software for AI for Science). 
\end{acks}